\documentclass[a4paper,12pt]{article}
\usepackage[utf8]{inputenc}
\usepackage{ae}
\usepackage{times,overcite}
\usepackage[pdftex]{color,graphicx}
\usepackage{geometry}

\begin{document}

\providecommand{\ignore}[1]{}
\newcommand{\comment}[1]{}
\renewcommand{\comment}[1]{{\color[rgb]{0,1,0}[#1]}}
\newcommand{\ket}[1]{{{|}{#1}\rangle}}
\newcommand{\bra}[1]{{\langle{#1}{|}}}
\newcommand{\braket}[2]{{\langle{#1}{|}{#2}\rangle}}

\title{Microwave quantum logic gates for trapped ions}

\author{C.~Ospelkaus, U.~Warring, Y.~Colombe, K.~R.~Brown,  J.~M.~Amini,\\ D.~Leibfried
 and D.~J.~Wineland}

\maketitle

\noindent

\textbf{
Control over physical systems at the quantum level is a goal shared by scientists in fields as diverse as metrology, information processing, simulation and chemistry. For trapped atomic ions, the quantized motional and internal degrees of freedom can be coherently manipulated with laser light\cite{bla08,lei03}. Similar control is difficult to achieve with radio frequency or microwave radiation because the essential coupling between internal degrees of freedom and motion requires significant field changes over the extent of the atoms' motion\cite{win98,lei03}. The field gradients are negligible at these frequencies for freely propagating fields; however,  stronger gradients can be generated in the near-field of microwave currents in structures smaller than the free-space wavelength\cite{osp08,chi08}. In the experiments reported here, we coherently manipulate the internal quantum states of the ions on time scales of 20 ns. We also generate entanglement between the internal degrees of freedom of two atoms with a gate operation\cite{sor99,sol99,mil00,osp08} suitable for general quantum computation\cite{bar95}. We implement both operations through the magnetic fields from microwave currents in electrodes that are integrated into a micro-fabricated trap structure and create an entangled state with fidelity 0.76(3). This approach, where the quantum control mechanism is integrated into the trapping device in a scalable manner, can potentially benefit quantum information processing\cite{osp08}, simulation\cite{chi08,sch09} and spectroscopy\cite{win98,sch05}.
}

\newpage
The quantized mechanical motion and internal states of trapped atoms can be coherently controlled by laser radiation. This optical control has led to the creation of nonclassical motional states~\cite{bla08,win98,lei03}, multiparticle entanglement~\cite{bla08}, scalable quantum logic techniques~\cite{spe09}, quantum simulation studies~\cite{por04,fri08,kim10} and quantum logic spectroscopy~\cite{sch05}.  The internal-to-motional state coupling is provided by the gradient of the driving field over the extent $\Delta x$ of the atom's motion \cite{win98,lei03}. For trapped atomic ions in their lowest motional quantum levels, $\Delta x \simeq 10\,\mathrm{nm}$, which is a significant fraction of an optical wavelength, resulting in an efficient coupling between motional and internal states. A microwave field propagating in free space exhibits very small variations of the field over $\Delta x$, and the corresponding coupling between motional and internal states is highly suppressed.

Notwithstanding the success of laser-based control, it is desirable to develop a microwave approach with equivalent capabilities  because of the relative ease of generating and controlling these fields. One method utilizes a combination of uniform microwave fields and static state-dependent potentials to provide internal-state to motion coupling~\cite{min01,cia03,joh09,foe09} and individual addressing~\cite{joh09,wan09}. Related to this are magnetic microwave or RF dressed state potentials for neutral atoms~\cite{for07}. Here, we implement the approach outlined in Ref.~\citeonline{osp08}, where the magnetic field from oscillating microwave currents in trap electrodes provides coherent internal state manipulation, and the gradient of this field leads to motional-internal state coupling. These interactions suffice for universal quantum information processing, because any unitary operation on qubits can be composed of a sequence of single-qubit rotations and a suitable entangling interaction~\cite{bar95}. Compared to laser-based approaches, microwave control has the potential to significantly improve the fidelity of operations due to its reduced sensitivity to ``spectator modes'' of ion motion~\cite{win98}, better control of field amplitudes and phases, and absence of spontaneous-emission decoherence~\cite{min01,osp08}. Furthermore, microwave near-field control could be incorporated in a chip-level library of transport, junction, storage and microwave manipulation components~\cite{ami10} to advance the integration of quantum control in scalable quantum information processing or simulation.

\begin{figure}[h]
  \centering
 \includegraphics[width=\columnwidth]{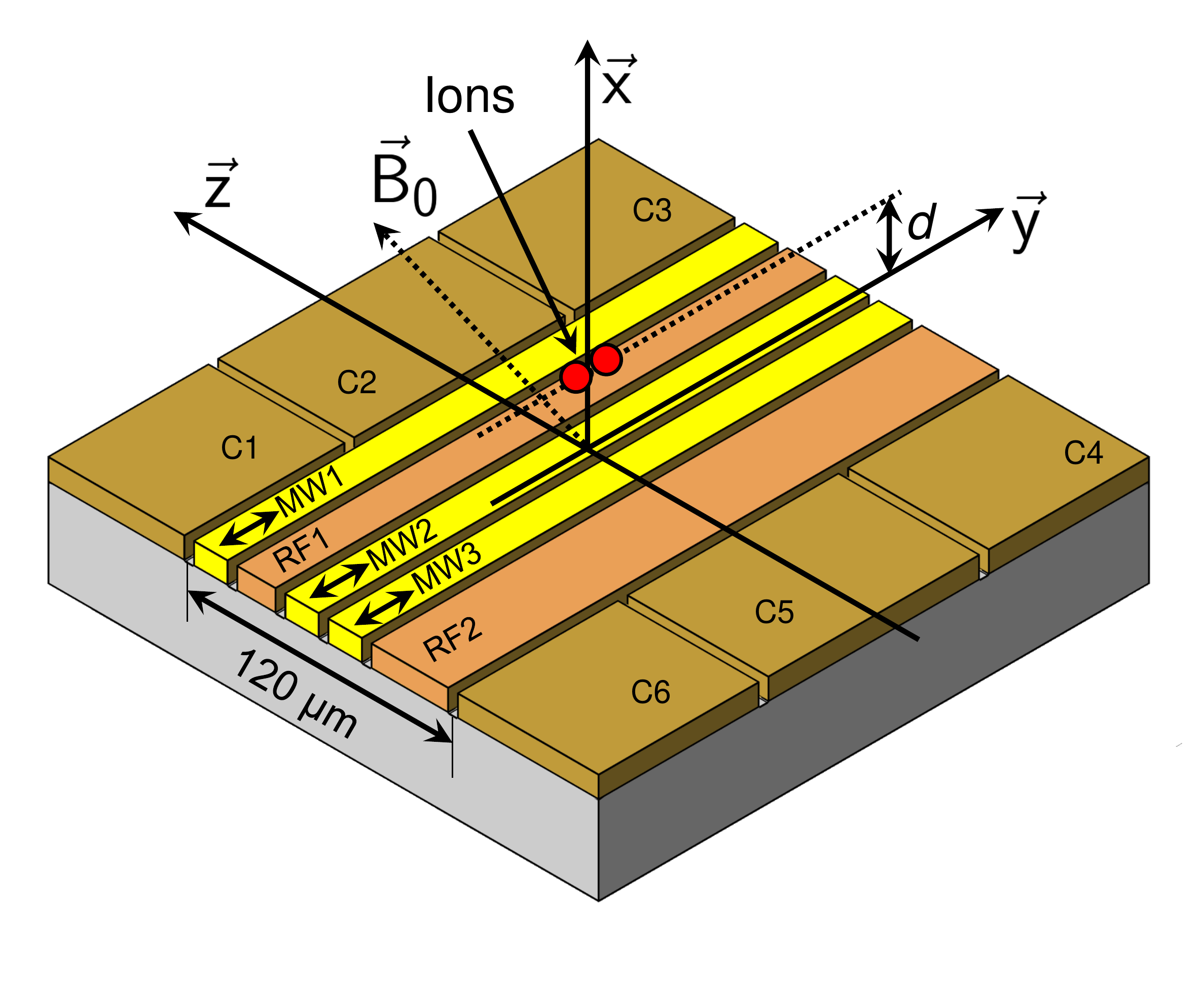}
  \caption{Central portion of the surface-electrode trap. The control electrodes are labeled as C1 to C6, trap-RF-electrodes as RF1 and RF2 and microwave lines as MW1 to MW3. The ions are held at a distance $d=30\,\mathrm{\mu m}$ above the surface. Compared to the symmetric trap structure of Ref.~\citeonline{osp08}, the asymmetric geometry of the microwave and RF electrodes used here yields approximately equal microwave currents in all three electrodes for field nulling at the ions while obtaining a deeper trap.}
\label{fig:chip}
\end{figure}

\begin{figure}[tb]
  \centering
 \includegraphics[width=10 cm]{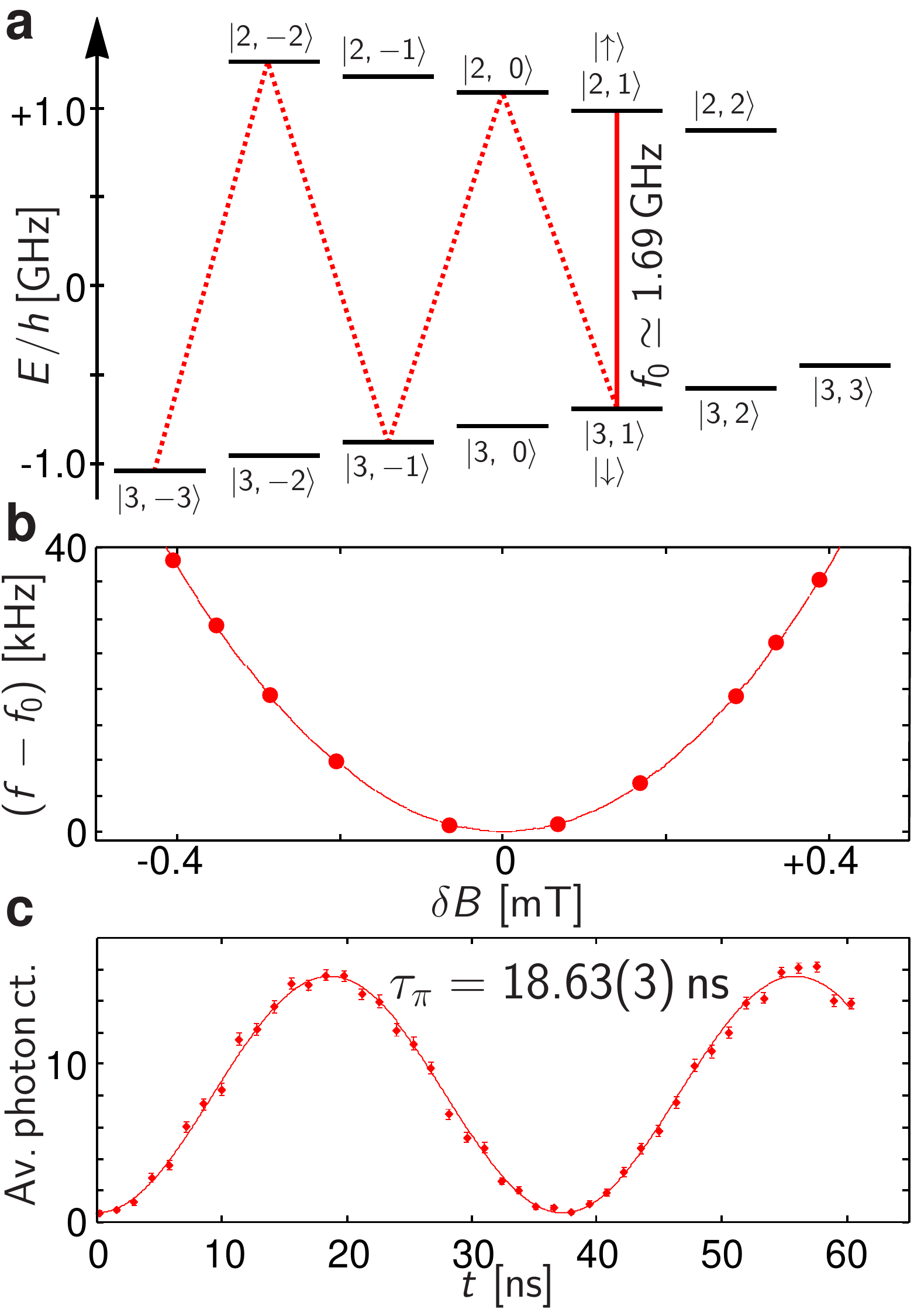}
  \caption{Level scheme of $^{25}\mathrm{Mg}^+$ (nuclear spin $I=5/2$) and spectroscopy of the qubit transition.
  (a) Ground-state hyperfine structure of $^{25}\mathrm{Mg}^+$ at $B_0\simeq21.3\,\mathrm{mT}$ and the microwave transitions used in the experiment.
  (b) Frequency of the $\ket{\downarrow}\leftrightarrow\ket{\uparrow}$ transition as a function of bias field detuning $\delta B$ from the field-independent point $B_0$. The standard errors of the measured frequencies are smaller than $5\,\mathrm{Hz}$.
  (c) Rabi flopping on the $\ket{\downarrow}\leftrightarrow\ket{\uparrow}$ transition induced by microwave current in MW 2, the electrode in closest proximity to the ion(s). Error bars represent the standard error of the mean (s.e.m.) for the average photon count in $400\,\mathrm{\mu s}$.
  }
\label{fig:fiqubit}
\end{figure}

The apparatus comprises a room-temperature surface-electrode ion trap~\cite{sei06} with 10 $\mu$m thick gold electrodes separated by gaps of 4.5 $\mu$m, deposited onto an insulating substrate (Fig.~\ref{fig:chip}). An oscillating potential (amplitude 25 V to 50 V, frequency $f_\mathrm{RF}=70.97\,\mathrm{MHz}$), applied to the RF electrodes in Fig.~\ref{fig:chip}a, provides pseudopotential confinement of $^{25}$Mg$^+$ ions in the radial ($x$ and $z$) directions at  a distance of 30 $\mu$m from the electrode surface (all other electrodes are held at RF ground). Along the trap $y$ axis, ions are confined with static potentials applied to control electrodes C1, C3, C4 and C6. The radial field resulting from these potentials is compensated by suitable potentials applied to electrodes C2 and C5. Single ion oscillation frequencies along $y$ can be adjusted between $300\,\mathrm{kHz}$ and $2\,\mathrm{MHz}$, and radial frequencies between $4\,\mathrm{MHz}$ and $10\,\mathrm{MHz}$. Microwave electrodes MW1, MW2 and MW3 support currents of order $100\,\mathrm{mA}$ to $1\,\mathrm{A}$ that produce an oscillating magnetic field $\vec{B}_\mathrm{osc}$ above the surface and are used to implement quantum control. To minimize thermal effects during these microwave pulses, we chose AlN as the electrode substrate. It provides a strong thermal link between the gold electrodes and a solid copper support via a thin layer of vacuum compatible epoxy. A static magnetic field $\vec B_0$ parallel to the trap surface and at an angle of $15^\circ$ with respect to the $z$ axis provides the internal state quantization axis. Superimposed $\sigma^-$-polarized laser beams nearly resonant with $3^2S_{1/2} \rightarrow 3^2P_{3/2}$ transitions and propagating parallel to $\vec B_0$ are used for optical pumping, Doppler laser cooling, and state detection through resonance fluorescence. Optical pumping prepares ions in the $\ket{F=3,m_F=-3}$ hyperfine ground state. Using microwave pulses resonant with hyperfine transitions, we can prepare an arbitrary pure state within the ground state manifold (see Fig.~\ref{fig:fiqubit}a) with fidelity higher than $0.98$ (see Methods). Furthermore, we can detect the population in any one of these states by applying a series of microwave $\pi$-pulses to transfer its population to $\ket{3,-3}$ and then detecting resonance fluorescence on the $\ket{3,-3}\rightarrow \ket{3^2P_{3/2},F=4,m_F=-4}$ optical cycling transition.

At a bias field $B_0\simeq21.3\,\mathrm{mT}$, we realize a first-order magnetic-field-independent transition between the qubit states $\ket{3,1}\equiv \ket{\downarrow}$ and $\ket{2,1}\equiv\ket{\uparrow}$ with ``carrier'' frequency $f_0\simeq1.69\,\mathrm{GHz}$ (solid line in Fig.~\ref{fig:fiqubit}a). To prepare $\ket{\downarrow}$, after optically pumping the ion(s) to $\ket{3,-3}$, we apply four microwave $\pi$-pulses (dotted lines in Fig.~\ref{fig:fiqubit}a). Fig.~\ref{fig:fiqubit}b shows Ramsey spectroscopy measurements of the $\ket{\downarrow}\rightarrow\ket{\uparrow}$ transition frequency as a function of magnetic field detuning $\delta B$ from $B_0$. At $\delta B=0$, the transition frequency depends on $\delta B$ only in second order, a feature that has been shown to enable long coherence times ($\approx 10\,\mathrm{s}$) despite moderate ambient magnetic field noise~\cite{lan05}.

Rabi flopping on the $\ket{\downarrow}\rightarrow\ket{\uparrow}$ transition with a $\pi$-time of $18.63(3)\,\mathrm{ns}$ (Fig.~\ref{fig:fiqubit}c) demonstrates the speed of single-qubit operations. This is more than 200 times shorter than a typical $\pi$-time ($4\,\mathrm{\mu s}$) obtained with continuous-wave Raman laser beams~\cite{jos09}. Moreover, it is nine orders of magnitude shorter than the coherence time achieved on a similar field-independent transition~\cite{lan05}. (For very short $\pi$-pulses realized with pulsed lasers, see Ref.~\citeonline{cam10}.)

To achieve coupling between motional and internal states, we apply sufficiently large oscillating magnetic field gradients at frequencies $f_\mathrm{s} \simeq f_0 \pm f_\mathrm{r}$ near the motional sidebands of the qubit transition carrier frequency $f_0$ (where $f_\mathrm{r}$ is a radial motional mode frequency). At the same time, we want to suppress $\vec{B}_\mathrm{osc}$ at the ion because it can cause undesirable off-resonant carrier excitation and AC-Zeeman shifts, analogous to AC-Stark shifts for optical fields~\cite{osp08}. In the geometry of Fig.~\ref{fig:chip}, there is a single combination of relative current amplitudes and phases in electrodes MW1, MW2 and MW3 that provides an oscillating magnetic field gradient at the ion without an oscillating magnetic field.

We find this combination by adjusting amplitudes and phases of the currents to minimize the AC-Zeeman shifts on the qubit and neighboring hyperfine transitions. To characterize the resulting field, we map the $xz$ spatial dependence of the AC-Zeeman shift on a selected magnetic dipole hyperfine-transition by displacing the ion radially from its nominal position with adjustments of the static trap potentials. Within a small region around the oscillating field zero, we expect $\vec B_\mathrm{osc}$ to be a quadrupole characterized by a gradient strength $B^\prime$ and an angle $\alpha$ (the angle of one principal axis of the quadrupole with respect to the $x$-axis; see Methods). A fit based on this model and to the experimental AC-Zeeman shift data yields $B^\prime=35.3(4)\,\mathrm{T/m}$ and $\alpha$ = 26.6(7)$^\circ$. For comparison, a plane-wave microwave field with the amplitude required to produce the $\pi$-time of Fig.~\ref{fig:fiqubit}c ($1.9\,\mathrm{mT}$), exhibits a gradient of only $0.068\,\mathrm{T/m}$. Knowledge of the orientation and gradient of the field can be used to predict sideband Rabi rates and to optimize the overlap with one of the radial motional modes. From the measured gradient and the assumption of optimal alignment of the mode with the gradient we calculate a sideband $\pi$-time~\cite{osp08} of approximately $190$~$\mu$s at a motional frequency of $6.5$~MHz for a single ion in its motional ground state (see Methods).

\begin{figure*}[t]
  \centering
 \includegraphics[width=12 cm,type=pdf,ext=.pdf,read=.pdf]{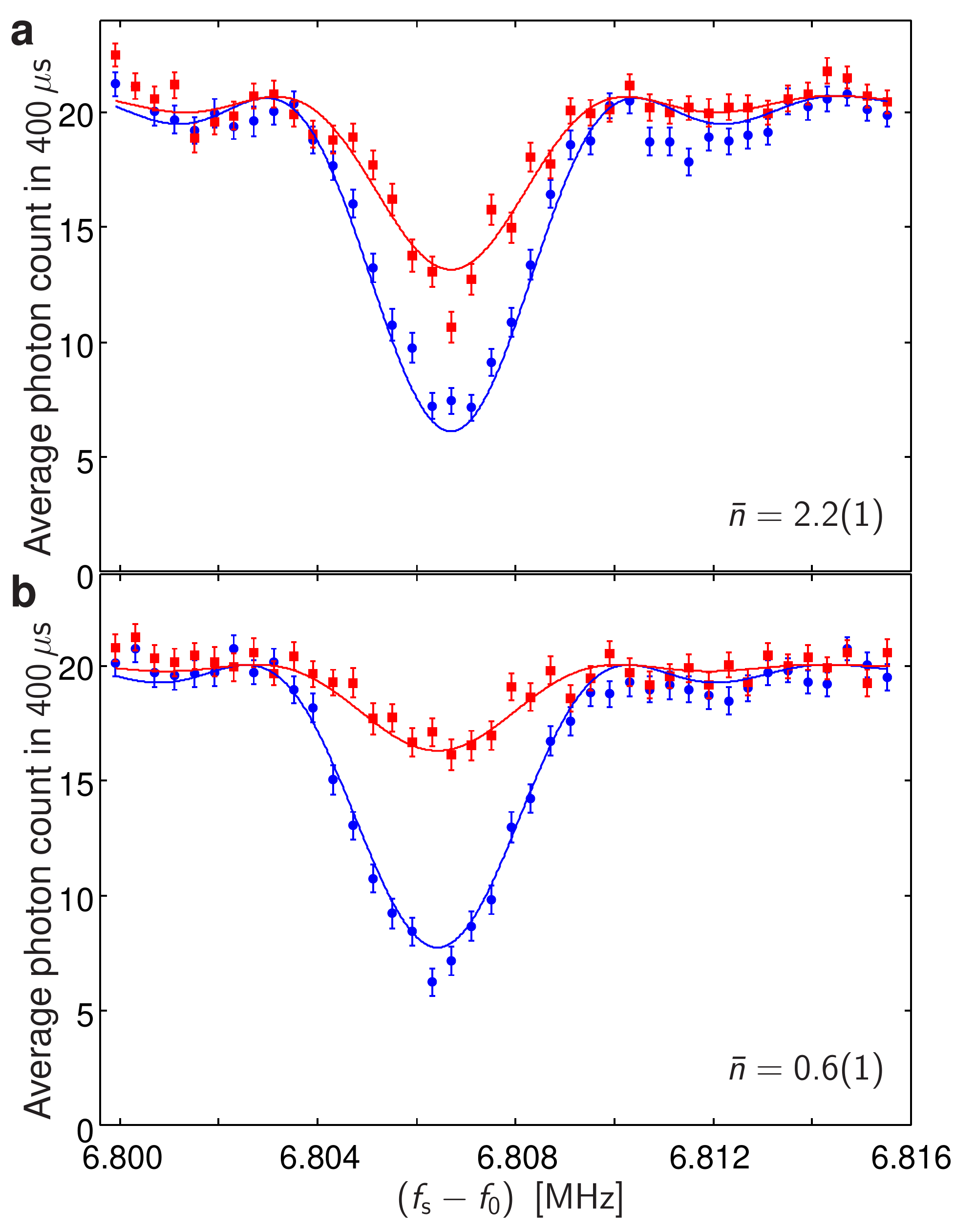}
  \caption{
  Microwave-motional sideband transitions.
  (a) Motion-subtracting $\ket{\uparrow \uparrow,n}\rightarrow \{\ket{\uparrow \downarrow,n-1},\ket{\downarrow \uparrow,n-1},\ket{\downarrow \downarrow,n-2}\}$ (red squares) and motion-adding $\ket{\downarrow \downarrow,n}\rightarrow \{\ket{\uparrow \downarrow,n+1},\ket{\downarrow \uparrow,n+1},\ket{\uparrow \uparrow,n+2}\}$ (blue circles) sideband transitions for a Doppler cooled two-ion radial rocking mode.
  (b) Sideband transitions after four sideband cooling cycles. Error bars represent the standard error of the mean (s.e.m.) for the average photon count in $400\,\mathrm{\mu s}$.
  }
\label{fig:sidebands}
\end{figure*}

After approximately optimizing the field configuration we implemented motional sidebands on one and two ions. As an example of a two-ion sideband application, we perform the following experiments on the selected radial rocking mode. First two ions are Doppler cooled and prepared in the $\ket{\uparrow \uparrow}$ state. We then apply a qubit $\pi$-pulse followed by a sideband pulse at $f_\mathrm{s}=f_0+f_\mathrm{r}$ (here $f_\mathrm{r}\approx 6.8\,\mathrm{MHz}$ for the chosen rocking mode). The sideband pulse induces transitions from $\ket{\downarrow \downarrow,n}$ to $\ket{\uparrow \downarrow,n+1}$, $\ket{\downarrow \uparrow ,n+1}$ and $\ket{\uparrow \uparrow ,n+2}$ where $n$ denotes the motional quantum number~\cite{kin98}.  Populations in $\ket{\downarrow}$ are subsequently transferred to $\ket{3,-3}$ and detected by laser fluorescence. The blue circles in Fig.~\ref{fig:sidebands}a show the result of a frequency scan of the sideband pulse. We now repeat the above experiment, but swap the order of the sideband pulse and the carrier $\pi$-pulse (red squares in Fig.~\ref{fig:sidebands}a). In this case, the sideband pulse drives transitions from $\ket{\uparrow \uparrow,n}$ to $\ket{\downarrow \uparrow,n-1}$, $\ket{\uparrow \downarrow ,n-1}$ and $\ket{\downarrow \downarrow ,n-2}$. If we assume a thermal motional state  (valid for Doppler cooling), a fit to the data gives $\bar{n}=2.2(1)$.

Multiple sideband cooling cycles can be used to cool a selected motional mode to near the ground state~\cite{lei03,kin98}. One sideband cooling cycle consists of transitions from $\ket{\uparrow \uparrow,n}$ to $\ket{\downarrow \uparrow,n-1}$, $\ket{\uparrow \downarrow ,n-1}$ and $\ket{\downarrow \downarrow ,n-2}$ where the pulse duration ($250~\mu$s) is optimized for maximal sideband asymmetry. This is followed by a repumping laser pulse resonant with the $\ket{\downarrow} \rightarrow \ket{3^2P_{3/2},m_J=-3/2,m_I=3/2}$ transition. From the relevant Clebsch-Gordan coefficients, we predict a 97~\% repumping efficiency to $\ket{\uparrow}$ after scattering 20 photons on average. Fig.~\ref{fig:sidebands}b shows a frequency scan of the motion-adding and motion-subtracting sidebands after four sideband cooling cycles. Assuming an approximate thermal distribution after cooling~\cite{kin98}, we obtain $\bar{n}=0.6(1)$, confirming the sideband cooling. Lower temperatures could not be achieved by adding more cooling cycles and the final value of $\bar{n}$ was most likely limited by the repumping photon recoil and by heating from ambient electric field noise~\cite{kin98} during the course of the sequence. Separately, we measured that the relevant motional heating rate was as low as $d \bar{n}/dt  = 0.2~\mathrm{ms}^{-1}$ at $f_{r}=7.7$~MHz; however, higher heating rates were also observed. We did not detect additional heating induced by the MW pulses.

\begin{figure*}[t]
  \centering
 \includegraphics[width=11 cm,type=pdf,ext=.pdf,read=.pdf]{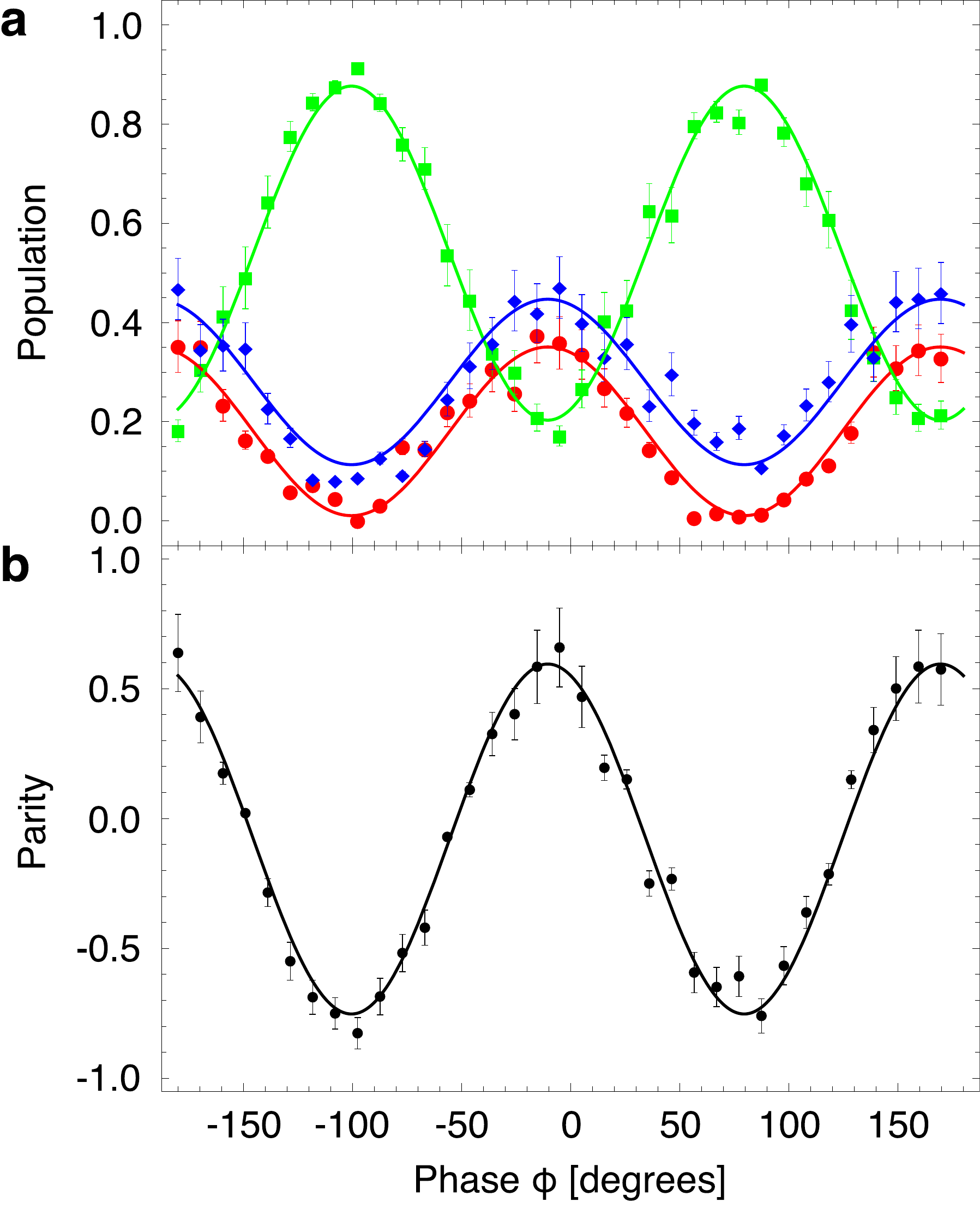}
  \caption{
  Populations and parity of the entangled state. After the entanglement operation is applied to the   $\ket{\downarrow \downarrow}$ state, an analysis $\pi/2$-pulse with relative phase $\phi$ is applied. Ideally, as $\phi$ is varied, $\Psi$ is transformed to superpositions of $\Psi$ and $\textstyle{\frac{1}{\sqrt{2}}} (\ket{\uparrow \downarrow}+\ket{\downarrow \uparrow})$, while parity and populations of the two ions oscillate at $\cos (2\phi+\phi_0)$. (a) Populations $P_{\ket{\uparrow \uparrow}}$ (red disks), $P_{\ket{\downarrow \downarrow}}$ (blue diamonds) and $(P_{\ket{\uparrow \downarrow}}+P_{\ket{\downarrow \uparrow}}$) (green squares). (b) Parity $\Pi=(P_{\ket{\uparrow \uparrow}}+P_{\ket{\downarrow \downarrow}})-(P_{\ket{\uparrow \downarrow}}+P_{\ket{\downarrow \uparrow}})$. Each data point represents an average of 300 experiments. We determine a fidelity of the entangled state $F=0.76(3)$ from the populations in (a). Error bars represent the standard error of the mean (s.e.m.) for the populations and parity.
  }
\label{fig:poppar}
\end{figure*}

Simultaneous application of two field gradients at $f_0 \pm (f_\mathrm{r}+\delta)$, with $\delta\ll f_\mathrm{r}$, is used to implement an entangling two-qubit gate~\cite{osp08}. Compared to light fields, derivatives of the microwave fields of higher order than linear have negligible effect on the ions~\cite{osp08}, so the ions are only Doppler cooled and prepared in the $\ket{\downarrow \downarrow}$ state. We then apply the two oscillating field gradients with $f_r= 7.6$~MHz (rocking mode) and $\delta$ = 4.9 kHz. To suppress decoherence caused by motional frequency instability, we apply the fields in two pulses of duration 200 $\mu$s, reversing the phase of one field in the second pulse~\cite{hay11}. For properly chosen pulse durations, the internal states ideally end in the entangled state $\ket{\Psi}= \textstyle{\frac{1}{\sqrt{2}}}(\ket{\downarrow \downarrow}-i \ket{\uparrow \uparrow})$ with the motion restored to its state before the gate~\cite{sor99}. Pulse imperfections and decoherence during the operation produce a mixed state with density matrix $\rho$ (after tracing over the motion). We characterize the fidelity of $\rho$ with respect to the ideal outcome, $F=\bra{\Psi}\rho \ket{\Psi} = 1/2(\rho_{\uparrow \uparrow, \uparrow \uparrow}+\rho_{\downarrow \downarrow, \downarrow \downarrow})+|\rho_{\uparrow \uparrow, \downarrow \downarrow}|$, by observing population oscillations as we sweep the phase $\phi$ of a $\pi/2$-pulse applied to the qubits after creating the entangled state~\cite{sac00}. The populations $\rho_{\downarrow \downarrow, \downarrow \downarrow}$, $(\rho_{\uparrow \downarrow,\uparrow \downarrow}+\rho_{\downarrow \uparrow,\downarrow \uparrow})$ and $\rho_{\uparrow \uparrow, \uparrow \uparrow}$ can be directly determined by fitting histograms for two, one or zero ions fluorescing to the histograms of the total fluorescence signal obtained during the sweep (see Methods). The resulting populations are shown in Fig.~\ref{fig:poppar}a. The magnitude of the density-matrix element $\rho_{\uparrow \uparrow, \downarrow \downarrow}$ is equal to the amplitude $A_{\Pi}$ of the parity $\Pi=(P_{\ket{\uparrow \uparrow}}+P_{\ket{\downarrow \downarrow}})-(P_{\ket{\uparrow \downarrow}}+P_{\ket{\downarrow \uparrow}})$ that oscillates as $A_\Pi \cos(2 \phi + \phi_0)$ as shown in Fig.~\ref{fig:poppar}b~\cite{sac00}.
By fitting sinusoids to the observed populations we extract a fidelity of $F= 0.76(3)$ with respect to the state $\ket{\Psi}$.

For scalable implementations of quantum information processing, the quality of all operations demonstrated here, in particular that of the entangling operation, must be improved considerably. This will require significant technical improvements; however, there are no apparent fundamental limits to fidelity. The most important factors currently limiting two-qubit gate fidelity are the stability of motional frequencies in the trap, and the precision with which we can suppress $\vec{B}_\mathrm{osc}$ at the positions of the ions. Fluctuations of the motional frequencies are on the order of 1 kHz during the gate, a sizeable fraction of the gate detuning $\delta$, and appear to be primarily caused by time-varying stray potentials on the trap-chip. It should be possible to suppress such stray fields by improving the surface quality of the electrodes, by reducing the amount of nearby dielectric materials (dust, residue of materials used during fabrication), with more careful cleaning of the electrode surfaces and by minimizing the exposure of the electrodes to UV light, which can generate charge via photoemission. We are able to suppress the oscillating field at the position of the ions to fluctuations with an amplitude of approximately $10~\mu$T, implying an AC-Zeeman shift smaller than 1 kHz. These fluctuations could be further suppressed by improving the amplitude stability and better shaping of the microwave pulses and by more sophisticated designs of the microwave current leads.

The presence of many processing zones on the same chip can introduce cross-talk between zones. Cross-talk for two-qubit gates can be suppressed by choosing substantially different mode frequencies for qubits in nearby "spectator" trap zones. For single-qubit rotations, it should be possible to null the field in spectator zones by applying local compensating fields. Composite pulse techniques can also prove useful to enhance spectator zone isolation~\cite{lev86}. Traps smaller than those used here should lead to faster operations~\cite{osp08}, provided that anomalous motional heating can be suppressed. Multi-ion operations similar to those demonstrated here might also be exploited for quantum simulation~\cite{chi08,sch09}, possibly with less stringent requirements on suppression of cross-talk.  For particles lacking the electronic structure needed for laser manipulation and readout, microwave sidebands would allow excitation of the motion conditioned on the internal state magnetic moment~\cite{hei90}. The presence of the motional excitation could then be observed with an ancilla ion, either via a shared motional mode~\cite{sch05}, or through the Coulomb interaction between ions held in separate traps~\cite{win98,hei90,bro11,har11}. In the latter case, this technique might enable a comparison of the proton and antiproton magnetic moment~\cite{win98,hei90}.

\section*{Methods Summary}

 We describe the oscillating magnetic field geometry in the vicinity of the ion(s) and the experimental determination of this geometry by mapping out AC-Zeeman shifts experienced by the ion as a function of its position. We also explain in more detail how state populations are extracted from fluorescence measurements to derive the fidelity of the entangled state produced by the two-qubit gate.

\section*{Methods}

\subsection*{Determination of the local microwave field}
Within a small region $\{\delta x, \delta y, \delta z \}$ around the oscillating microwave field null point, we expect $\vec B_\mathrm{osc}$ to be a quadrupole,
\begin{eqnarray}
\label{eq:oscbfield}
\vec B_\mathrm{osc}=B^\prime \cos(2\pi f_\mathrm{s} t)
 \left(
    \begin{array}{ccc}
      \cos( 2\alpha) & 0 & \sin( 2\alpha) \\
      0           & 0 & 0           \\
      \sin(2 \alpha) & 0 &  -\cos( 2\alpha)
    \end{array}
  \right)
  \cdot
  \left(
    \begin{array}{c}
      \delta x\\
      \delta y\\
      \delta z
    \end{array}
  \right), \nonumber
\end{eqnarray}
in the coordinate system of Fig.~\ref{fig:chip}, where $B^\prime$ and $\alpha$ (the angle of one principal axis of the quadrupole with respect to the $x$-direction in Fig.~\ref{fig:chip}) characterize the strength and orientation of the quadrupole. We map the $xz$ spatial dependence of the AC-Zeeman shift on the magnetic dipole transition $\ket{3,1}\rightarrow \ket{2,0}$  by displacing the ion radially from its nominal position with adjustments of the control electrode potentials that are derived from simulations. We determine the AC-Zeeman shifts for a set of radial displacements $\{\delta x,\delta z\}$ and extract $B^\prime=35.3(4)\,\mathrm{T/m}$ and $\alpha$ = 26.6(7)$^\circ$ from a fit to this data. As a consistency check we probe the relevant one-ion sideband with frequency near $6.5$~MHz and find a $\pi$-time for the ion in its motional ground state of approximately $260~\mu$s (extracted by a model fit). The  discrepancy of $27~\%$ between the observed $\pi$-time and that inferred from the measured gradient is most likely due to a misalignment between the maximal gradient direction and the axis of the interrogated motional mode (To maintain a reasonable trap depth, the motional mode could not be perfectly overlapped with the maximal gradient).

\subsection*{Determination of populations from state-dependent fluorescence}
During one detection period (duration 100 $\mu$s) we typically detect on average $\approx 0.3$ counts if both ions are projected into $\ket{ \uparrow}$, and approximately 11 additional counts for each ion in state $\ket{ \downarrow}$.

Reference histograms were derived by recording photon counts for an on-resonance Ramsey experiment with two ions where the phase $\theta$ of the second $\pi/2$-pulse was varied. We established in independent experiments that the net effect of transferring in and out of the qubit basis, and the Ramsey pulses, gave population errors not exceeding $0.02$.  In that case we assume the probability of observing zero, one or two ions bright is:
\begin{eqnarray}
P_0(\theta)=\cos^4(\theta/2) \nonumber \\
P_1(\theta)=\sin^2(\theta)/2 \nonumber \\
P_2(\theta)=\sin^4(\theta/2).
\end{eqnarray}
We model each histogram for two, one or zero ions bright respectively, by a sum of three Poissonian distributions with variable weights and average counts. The weights and averages are determined by a simultaneous least-squares fit to the experimentally determined histograms for all phases $\theta$. To reduce the number of free parameters we fixed the two-ion bright average to be two times the one-ion bright average. With this choice, the fit contains eight free parameters and has a reduced $\chi^2$-value of 1.34. Using the histograms representing two, one and zero ions bright, we do a second least squares fit to the histograms obtained while scanning the phase $\phi$ of the $\pi/2$-pulse following the entangling gate. Each population is fit to functions $a_{k} \cos (2\phi+\phi_0)+a_{0,k}$ with $k~ \epsilon~ \{{\rm two~bright},~ {\rm one~bright},~{\rm zero~bright}\} $ and a common phase offset $\phi_0$. The least-squares fits are shown in Fig.~\ref{fig:poppar}a. These fits directly yield the parity amplitude $A_\Pi$, the phase offset $\phi_0$, and the offset from zero shown in Fig.~\ref{fig:poppar}b.

The final uncertainty in fidelity $F$ is derived from the uncertainties in the populations fit. In addition to the data analysis described above, we applied several other parameterizations of the histograms to fit to the Ramsey-reference experiment. In all instances the resulting gate fidelities agreed with our stated result within the respective uncertainties.

\section*{Acknowledgments}

We thank M.~J.~Biercuk, J.~J.~Bollinger and A.~P.~VanDevender for experimental assistance, J. C. Bergquist, J. Chou and T. Rosenband for the loan of a fiber laser, R. Jordens and E. Knill for comments on the manuscript and D. Hanneke and J. Home for discussions. We thank P. Treutlein for discussions on micro-fabrication techniques. This work was supported by IARPA, ONR, DARPA, NSA, Sandia National Laboratories and the NIST Quantum Information Program. This paper, a submission of NIST, is not subject to US~copyright.

\section*{Author Contributions}
C.O. participated in the design of the experiment and built the experimental apparatus, collected data, analyzed results and wrote the manuscript. U.W. participated in building the experimental apparatus, collected data and analyzed results. Y.C. developed chip fabrication methods and fabricated the ion trap chip. K.R.B. participated in the design of the experiment, developed chip
fabrication methods, and helped build parts of the experiment. J.M.A. developed chip fabrication methods and automated experiment control and data taking. D.L. participated in the design of the experiment, collected data and maintained laser systems. D.J.W. participated in the design and analysis of the experiment. All authors discussed the results and the text of the manuscript.

\section*{Author Information}

Reprints and permissions information is available at www.nature.com/reprints. The authors declare no competing financial interests. Readers are welcome to comment on the online version of this article at
www.nature.com/nature. Correspondence and requests for materials should be addressed to
C.O. (christian.ospelkaus@iqo.uni-hannover.de).

\newpage

\end{document}